\documentclass[aps,pra,twocolumn,superscriptaddress]{revtex4}
\usepackage{amsfonts}
\usepackage{amsmath}
\usepackage{amssymb}
\usepackage{graphicx}

\begin{document}

\title{\small Dynamics of rotating spin-orbit-coupled Bose-Einstein condensates in a quasicrystalline optical lattice}
\author{\footnotesize Qingbo Wang}
\affiliation{\footnotesize Key Laboratory for Microstructural Material Physics of Hebei Province, School of Science, Yanshan University, Qinhuangdao 066004, People's Republic of China}
\affiliation{\footnotesize School of Physics and Technology, Tangshan Normal University, Tangshan 063000, People's Republic of China}
\author{\footnotesize Jinguo Hu}
\affiliation{\footnotesize Key Laboratory for Microstructural Material Physics of Hebei Province, School of Science, Yanshan University, Qinhuangdao 066004, People's Republic of China}
\author{\footnotesize Xianghua Su}
\affiliation{\footnotesize Key Laboratory for Microstructural Material Physics of Hebei Province, School of Science, Yanshan University, Qinhuangdao 066004, People's Republic of China}
\author{\footnotesize Linghua Wen}
\email{linghuawen@ysu.edu.cn}
\affiliation{\footnotesize Key Laboratory for Microstructural Material Physics of Hebei Province, School of Science, Yanshan University, Qinhuangdao 066004, People's Republic of China}

\begin{abstract}
\footnotesize
We investigate the dynamics of rotating pseudo-spin-1/2 Bose-Einstein condensates (BECs) with Rashba spin-orbit coupling (SOC) in a quasicrystalline optical lattice (QOL). For given parameters, the system evolves from an initial heliciform-stripe phase into a final visible vortex necklace with a giant vortex and a hidden vortex necklace. Simultaneously, the corresponding spin texture undergoes a transition from a meron-antimeron pair to a half-antiskyrmion necklace. During the dynamic evolution process, the angular momentum increases gradually, and then approaches to a convergent value. Furthermore, typical quantum phases of rotating two-component BECs with SOC in different external potentials are summarized.
\end{abstract}

\maketitle

\footnotesize
Spin-orbit coupling (SOC) effects in Bose-Einstein condensates (BECs) have been considered of major theoretical and experimental interest in the last decade \cite{Lin2011,Zhai2015}. In particular, the significant experimental developments\cite{Wu2016,Huang2016} on the realization of high-dimensional SOC has promoted the investigations of exotic quantum phases in rotating spin-orbit-coupled BECs \cite{Aftalion2013,Wang2017,Wang2020} in various trapping potentials. In fact, various trapping potentials may significantly affect the stationary states and dynamic behaviors of the BECs. Recently, a novel quasicrystalline optical lattice (QOL) with long-range order and high-rotational symmetry has become experimentally achievable \cite{Viebahn2019}. Matter-wave interference and Bloch oscillation are predicted for a BEC in the QOL \cite{Niu2020}. What interests us very much is the dynamic properties of rotating BECs with SOC in the QOL.

We consider a quasi-2D system of rotating pseudo-spin-1/2 BEC with Rashba SOC trapped in a QOL. In the mean-field framework, the dynamics can be well described by the dissipative coupled Gross-Pitaevskii (GP) equations \cite{Wen2010,Wen2012,Wen2014}
\begin{eqnarray}
(i-\gamma )\hbar \frac{\partial \psi _{1}}{\partial t} &=&\left[ -\frac{\hbar ^{2}\nabla ^{2}}{2m}+V_{\mathrm{Q}}(\mathbf{r})+g_{11}\left\vert \psi_{1}\right\vert ^{2}+g_{12}\left\vert \psi _{2}\right\vert ^{2}\right] \psi_{1}  \notag \\
&&-\Omega L_{z}\psi _{1}+\hbar \lambda (\partial _{x}-i\partial _{y})\psi_{2}, \\
(i-\gamma )\hbar \frac{\partial \psi _{2}}{\partial t} &=&\left[ -\frac{\hbar ^{2}\nabla ^{2}}{2m}+V_{\mathrm{Q}}(\mathbf{r})+g_{21}\left\vert \psi_{1}\right\vert ^{2}+g_{22}\left\vert \psi _{2}\right\vert ^{2}\right] \psi_{2}  \notag \\
&&-\Omega L_{z}\psi _{2}-\hbar \lambda (\partial _{x}+i\partial _{y})\psi_{1},
\end{eqnarray}
where $\psi _{1}$ and $\psi _{2}$ are the component wave functions, $m$ is the atomic mass, and $\gamma $ is the dissipation parameter. The phenomenological model is a variation of that in \cite{Tsubota2002} and a generalization of that of a rotating scalar BEC \cite{Wen2010,Wen2012}, and has good predictive power. In particular, this model enable one not only to find the steady state of a rotating system but also to study the whole dynamical process toward the final steady state. When the dissipation term is completely switched off, our simulation based on the time-dependent GP equations shows that there is no steady state. In addition, the motion of generated vortices remains turbulent and rather irregular, and there is no vortex lattice (or vortex necklace) formation. We take $\gamma =0.03$ throughout this work, corresponding to a temperature of about 0.1$T_{c}$ \cite{Wen2010,Wen2012}. In fact, the variation of nonzero $\gamma $ only influences the relaxation time scale but does not change the dynamics of the vortex formation and the ultimate steady structure of the rotating system. The initial wave functions are normalized as $\int [\left\vert \psi_{1}\right\vert ^{2}+\left\vert \psi _{2}\right\vert ^{2}]dxdy=N$ with $N$ being the initial number of atoms. The QOL is an eightfold symmetric optical lattice plus a harmonic trap, which can be defined by \cite{Viebahn2019}
\begin{equation}
V_{\mathrm{Q}}(\mathbf{r})=V_{0}\sum_{i=1}^{4}{\cos ^{2}(\frac{\mathbf{\hat{G}_{i}}}{2}\cdot \mathbf{r})}+\frac{1}{2}m{\omega _{\perp }^{2}}{\mathbf{r}}^{2},  \label{Potential}
\end{equation}
where $V_{0}$ is the lattice strength. The reciprocal lattice vectors are $\mathbf{\hat{G}_{1}}\propto (1,0)$, $\mathbf{\hat{G}_{2}}\propto (1,1)/\sqrt{2}$, $\mathbf{\hat{G}_{3}}\propto (0,1)$, and $\mathbf{\hat{G}_{4}}\propto(-1,1)/\sqrt{2}$ \cite{Niu2020}. $\omega _{\bot }$ is radial oscillation frequency, and $r=\sqrt{x^{2}+y^{2}}$. The interaction parameters are given by $g_{jj}=2\sqrt{2\pi }a_{jj}\hbar ^{2}/ma_{z}\;(j=1,2)$ and $g_{12}=g_{21}=2\sqrt{2\pi }a_{12}\hbar ^{2}/ma_{z}$, where $a_{jj}$ and $a_{12}$ are the s-wave scattering lengths between intra- and inter-component atoms, and $a_{z}=\sqrt{\hbar /m\omega _{z}}$ is the oscillation length in the $z$ direction. $\lambda $ is the strength of isotropic SOC. $\Omega $ is the rotation frequency along the z direction, and $L_{z}=i\hbar (y\partial _{x}-x\partial _{y})$ is the $z$ component of the angular momentum operator. The angular momentum is defined as $\langle L_{z}\rangle =i\hbar \int \mathbf{\Psi }^{\dagger }(\mathbf{r})(y\partial _{x}-x\partial _{y})\mathbf{\Psi }(\mathbf{r})d\mathbf{r}$ with $\mathbf{\Psi }(\mathbf{r})=(\psi _{1},\psi _{2})^{\mathrm{T}}$. The energy functional of the system is given by
\begin{eqnarray}
E&=\sum\limits_{j=1,2}\int\mathrm{d^2}\mathrm{r}\left\{\frac{\hbar^2}{2m}\vert\nabla\psi_j\vert^2+V_\mathrm{Q}\vert\psi_j\vert^2+\frac{g_{jj}}{2}\vert\psi_j\vert^4-\Omega\psi_j^*L_z\psi_j \right.  \notag \\ &\left.+(-1)^{3-j}h\lambda\psi_j^*[\partial_x+(-1)^ji\partial_y]\psi_{3-j}\right\}+\int g_{12}\vert\psi_1\vert^2\vert \psi_2\vert^2\mathrm{d^2}\mathrm{r}.
\end{eqnarray}

By introducing the notations $\widetilde{r}=r/a_{0}$, $a_{0}=\sqrt{\hbar/m\omega _{\perp }}$, $\widetilde{t}=\omega _{\perp }t$, $\widetilde{V}(r)=V_{Q}(r)/\hbar \omega _{\perp }$, $\widetilde{\Omega }=\Omega /\omega_{\perp }$, $\widetilde{L}_{z}=L_{z}/\hbar $, $\widetilde{\psi }_{j}=\psi_{j}a_{0}/\sqrt{N}(j=1,2)$, $\beta_{jj}=g_{jj}N/\hbar \omega_{\perp}a_{0}^{2}(j=1,2)$, and $\beta_{12}=\beta _{21}=g_{12}N/\hbar \omega_{\perp }a_{0}^{2}$, we obtain the dimensionless dissipative coupled GP equations,
\begin{eqnarray}
(i-\gamma )\partial _{t}\psi _{1} &=&\big( -\frac{1}{2}\nabla^{2}+V+\beta_{11}\left\vert \psi _{1}\right\vert ^{2}+\beta _{12}\left\vert\psi_{2}\right\vert ^{2}\big) \psi _{1}  \notag \\
&&-\Omega L_{z}\psi _{1}+\lambda (\partial _{x}-i\partial _{y})\psi _{2}, \\
(i-\gamma )\partial _{t}\psi _{2} &=&\big( -\frac{1}{2}\nabla^{2}+V+\beta_{12}\left\vert \psi _{1}\right\vert ^{2}+\beta _{22}\left\vert\psi_{2}\right\vert ^{2}\big) \psi _{2}  \notag \\
&&-\Omega L_{z}\psi _{2}-\lambda (\partial _{x}+i\partial _{y})\psi _{1},
\end{eqnarray}
where the tilde is omitted for simplicity. The spin texture is defined by $\mathbf{S}=\overline{\mathbf{\chi}}\mathbf{\sigma}\mathbf{\chi} $ with $\mathbf{\sigma}=(\sigma_x,\sigma_y,\sigma_z)$ being the Pauli matrix, where $\mathbf{\chi}=[\chi_1,\chi_2]^{\mathrm{T}}$ with $\chi_j=\psi_j/\sqrt{\rho}$
($j=1,2$) and $\rho=|\psi_1|^{2}+|\psi_2|^{2}$. The topological charge density can be written as $q(r)=\frac{1}{4\pi }\mathbf{S}\cdot(\frac{\partial \mathbf{S}}{\partial x}\times \frac{\partial \mathbf{S}}{\partial y})$, and the topological charge is given by $Q=\int q(r)dxdy$.

We numerically solve the GP equations with the split-step Fourier method \cite{Wen2010,Wen2012}. The initial quantum state can be obtained by using the imaginary-time propagation method \cite{Wen2010,Wen2012,Wen2014} for $\Omega=0$. We consider $^{87}$Rb atoms which are confined in a $(\omega_{\perp }, \omega_{z})=2\pi\times(10, 100)$ Hz trap, and the length scale is $a_0=3.411\mu$m. The numerical grids are chosen as $256\times256$, and the actual system is set as $20\times20{a_0}^2$. We introduce a ratio between inter- and intra-species interaction, i.e., $\delta =\beta _{12}/\beta _{11}$, and we set $\beta_{11}=\beta_{22}=200$ with $a_{11}=81.35a_B$($a_B$ being the Bohr radius).

We first prepare a spin-orbit-coupled BEC trapped in a stationary QOL. Figure 1 shows the typical dynamics of the component densities and phases after the QOL begin to rotate suddenly with $\Omega=0.6$. For a BEC with SOC in a stationary harmonic trap, the typical quantum phases are plane-wave phase and stripe phase \cite{Zhai2015}. Here the system initially exhibits spatially separated heliciform-stripe phase [Fig. 1(a)] which is resulted from the interplay of the SOC, QOL and the strong interspecies repulsion. Then the component densities are elongated along the $x$ axis or the $y$ axis, and the heliciform stripes are broken and separated into fragments [Fig. 1(b)]. When $t=5$, the system becomes irregular and reaches the minimum in the density distribution, and complex turbulent oscillations appear in the phase distribution [Fig. 1(c)], which makes the boundary of the system unstable and excites the surface waves propagating along the
surfaces. With further time evolution, the central surface waves develop into visible vortices and hidden vortices \cite{Wen2010,Wen2012}, and the boundary
ones become ghost vortices [Figs. 1(d)-1(e)]. The system finally forms a visible vortex necklace with a giant vortex, a hidden vortex necklace and numerous ghost vortices [Fig. 1(f)]. One of the reasons for the vortex necklace formation in Fig.1(f) is the presence of the QOL.
\begin{figure}[htpb]
\centerline{\includegraphics*[width=6.8 cm]{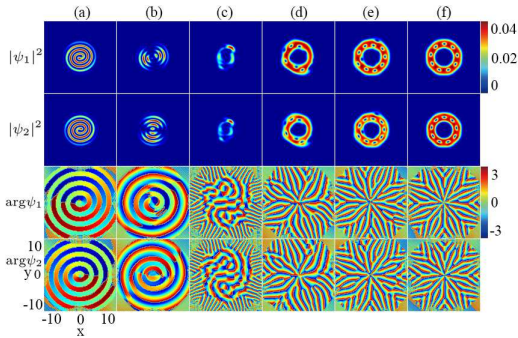}}
\caption{\scriptsize Time evolution of the density and phase distributions in the QOL with $V_0=2$ when the system suddenly begins to rotate with $\Omega=0.6$, where $\protect\lambda=3$, $\protect\delta=4$, and $\protect\gamma=0.03$. The time is (a) $t=0$, (b) $t=0.1$, (c) $t=5$, (d) $t=15$, (e) $t=50$, and (f) $t=300$. The length and time are in units of $a_0$ and $1/\protect\omega_{\perp}$, respectively.}
\end{figure}

In Fig. 2, we display the typical transitions of the topological charge density and spin texture. Our numerical calculation shows that the local topological charges in Figs. 2(a3) and 2(a4) approach $Q=0.5$ and $Q=-0.5$, which indicates that the initial-state spin texture in Fig. 2(a2) is a circular half-skyrmion-half-antiskyrmion (meron-antimeron) pair \cite{Liu2012}. From Figs. 2(b1)-2(b4), there are ten half-antiskyrmions with respective local topological charge $Q=-0.5$, and they seem to be forming a half-antiskyrmion necklace along a large circle. Since the transitional spin texture is unstable, we can not determine exactly what the topological defect is. Shown in Figs. 2(c1) and 2(c2) are the topological charge density and spin texture of the steady state, and the local amplifications are given in Figs. 2(c3) and 2(c4). The topological structure is a half-antiskyrmion necklace with respective local topological charge $Q=-0.5$. In short, the spin texture undergoes a transition from a meron-antimeron pair to a half-antiskyrmion necklace.
\begin{figure}[htbp]
\centerline{\includegraphics*[width=6 cm]{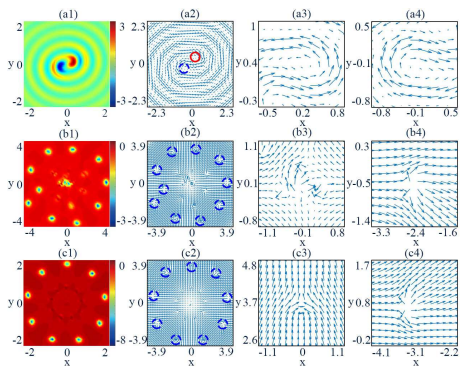}}
\caption{\scriptsize The transition of topological charge densities and spin textures in the QOL. Figs. 2(a1)-(a4), 2(b1)-(b4), and 2(c1)-(c4) correspond to Figs. 1(a), 1(e), and 1(f), respectively. The left two columns represent the topological charge densities and spin textures, the right two columns represent the local enlargements of spin textures, respectively. The red solid circle and blue dotted circle denote a half-skyrmion and a half-antiskyrmion, respectively. The unit length is $a_0$.}
\end{figure}

The dynamic process can also be characterized by the time evolution of the average angular momentum per atom $\langle L_z\rangle$, where the dependence of $\langle L_z\rangle$ on $\Omega$, $\lambda$, $\delta$, and $V_0$ are displayed in Fig. 3, respectively. Take the black solid curve as an example, when the system suddenly begins to rotate with $\Omega=0.6$, $\langle L_z\rangle$ increases rapidly with the time evolution ($0<t\leq15$), and then gradually ($15<t\leq100$) approaches a maximum value (equilibrium value). Physically, the combined effects of the continuous input of angular momentum, the quantum fluid nature, SOC, QOL and the dissipation lead to the formation of almost the same steady vortex structure in the two components (e.g., see Fig.1(f)).
\begin{figure}[htbp]
\centerline{\includegraphics*[width=6.5 cm]{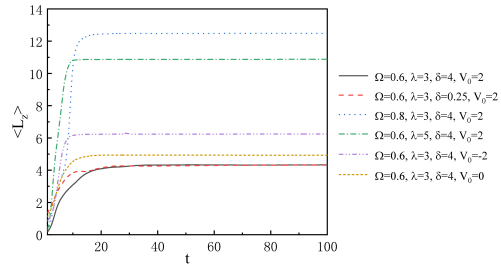}}
\caption{\scriptsize Time evolution of the average angular momentum per atom $\langle L_z\rangle$ with $\protect\gamma=0.03$. Here $\langle L_z\rangle$ and $t$ are in units of $\hbar$ and $1/\protect\omega_{\perp}$, respectively.}
\end{figure}

\begin{table}[!ht]
\caption{\footnotesize{Typical phases of rotating binary BECs with SOC}}
\label{table1}%
\centering
\begin{tabular}{p{83pt}p{155pt}}
\hline\hline
\footnotesize{External potential} & \footnotesize{Quantum phase ($\delta>1$)} \\ \hline
\scriptsize{Harmonic trap} & \scriptsize{Segregated symmetry preserving condensates with a giant skyrmion, stripes, vortex and peak lattices \cite{Aftalion2013}} \\
\scriptsize{Toroidal trap} & \scriptsize{Triangular vortex lattice, Anderson-Toulouse coreless vortex \cite{Wang2017}} \\
\scriptsize{Standard optical lattice} & \scriptsize{Square vortex lattice, vortex chain \cite{Yang2019}}\\
\scriptsize{Quasicrystalline optical lattice} & \scriptsize{Visible vortex necklace with a giant vortex and a hidden vortex necklace, vortex necklace with a bright soliton, triangular vortex lattice with symmetrical stripes, etc} \\ \hline\hline
\end{tabular}
\end{table}

In summary, the rotating pseudo-spin-1/2 BECs with Rashba SOC in a QOL can show interesting and unusual dynamic behaviors. For fixed parameters, the system gradually evolves from the initial heliciform-stripe phase to the steady visible vortex necklace with a giant vortex and a hidden vortex necklace. At the same time, the spin texture experiences a structural phase transition from a meron-antimeron pair into a half-antiskyrmion necklace. Furthermore, the temporal evolution of the angular momentum is revealed, where the angular momentum gradually increases to an equilibrium value. Finally, the typical quantum phases of rotating two-component BECs with SOC in different external potentials (including the different rotating results between a QOL and a standard OL) are briefly summarized in Table 1, where the ones without references are our simulation results. Our findings have provided new understanding for the physical properties of ultracold quantum gases.

\noindent{\textbf{Acknowledgments}}

\scriptsize{This work was partially carried out during a visit of the corresponding author (LW) to the research group of Prof. W Vincent Liu at The University of Pittsburgh. QW thanks Zhenxia Niu for helpful discussions. This work was supported by the National Natural Science Foundation of China (Grant Nos. 11475144 and 11047033), the Natural Science Foundation of Hebei Province (Grant Nos. A2019203049 and A2015203037), and Research Foundation of Yanshan University (Grant No. B846).}

\end{document}